\documentclass{aa}
\usepackage{amssymb,amsmath}
\usepackage{natbib}
\bibpunct{(}{)}{;}{a}{}{,}
\usepackage{graphicx}
\usepackage{txfonts}
\begin{document}

\title{X-ray variability patterns in blazars}

\author{K. Moraitis \and A. Mastichiadis}

\institute{Department of Physics, University of Athens, Panepistimiopolis, GR 15783 Zografou, Greece}

\date{Received .../ Accepted ...}

\abstract {}{We study the expected variability patterns of blazars within the two-zone acceleration model putting special emphasis on flare
shapes and spectral lags.} {We solve semi-analytically the kinetic equations which describe the particle evolution in the acceleration and
radiation zone. We then perturb the solutions by introducing Lorentzian variations in its key parameters and examine the flaring behavior of the
system. We apply the above to the X-ray observations of blazar \object{1ES\ 1218+304} which exhibited a hard lag behavior during a flaring
episode and discuss possibilities of producing it within the context of our model.} {The steady-state radio to X-rays emission of \object{1ES\
1218+304} can be reproduced with parameters which lie well within the ones generally accepted from blazar modeling. Additionally, we find that
the best way to explain its flaring behavior is by varying the rate of particles injected in the acceleration zone.}{}


\maketitle

\section{Introduction}

Blazars, a subclass of Active Galactic Nuclei, show variability which is observed across the electro-magnetic spectrum and can, in some cases,
be as fast as a few minutes \citep{aharonian07,albert07}. The working blazar scenario is that we observe their radiation coming from a jet which
is directed, within a small angle, to our line-of-sight. The observed spectrum, which is clearly non-thermal in origin, shows a characteristic
double-peak in a $\nu F_\nu$ plot, and is produced from a population of relativistic particles, presumably accelerated by shock waves within the
jet. While there are still open questions concerning the origin of the high energy gamma radiation, there is a consensus that the emission at
lower frequencies, which usually extends from radio to X-rays, comes from electron synchrotron radiation. The observed variability can then be
attributed directly to conditions in the acceleration region \citep[for a recent review of the various blazar models see][]{boettcher10}.

The detailed X-ray observations of many blazars reveal that these objects show a rich and complex structure in their flaring behavior.
Especially interesting are the trends in the lags between the soft and hard X-ray bands. The first observations which were detailed enough to
show time structure in various energy bands \citep{takahashietal96} revealed the blazar \object{Mrk 421} having a soft lag flare -- that is the
soft X-rays lagged behind the hard ones. This was explained as synchrotron cooling: lower energy electrons cool with a slower rate than higher
energy ones and therefore soft X-rays appear after the harder ones, i.e. soft lags hard.

Subsequent observations, however, showed cases of the opposite trend \citep{takahashietal2000}. This could not be explained by synchrotron
cooling, but it was attributed to particle acceleration. As fresh particles are accelerated to high energies, they radiate first soft and later
hard photons, i.e. hard lags soft.

Long, uninterrupted observations \citep{brinkmannetal05} revealed that \object{Mrk 421} shows both kinds of lags, with typical timescales of the
order of $10^3$s. The same mixed behavior was also verified for the same source by \citet{tramacere09}. Finally, blazar \object{1ES\ 1218+304}
showed a hard lag during a giant flare \citep{sato08}.

While several time-dependent models have been put forward to explain time variations \citep{mastikirk97,likusunose00,boettcherchiang2002}, hard
lag behavior needs a particle acceleration scheme to be taken explicitly into account \citep{kirima98,kusunoseetal00}. \citet[][hereafter
KRM]{kirima98} have shown that both soft and hard lags can be obtained during a flare once a detailed scheme for particle acceleration is
considered. Adopting a two-zone model in which particles are energized in an acceleration zone by the first-order Fermi process and then escape
and radiate in a radiation zone, they argued that flares produced by an increase in the number of the injected particles can produce soft lags
if the radiating electrons are in the energy regime where cooling is faster than acceleration and hard lags when the two timescales are
comparable.

More recently, two-zone models which include SSC losses have been developed in \citet{graff08,weidinger10}. In particular,
\citeauthor{weidinger10} include both first and second-order Fermi acceleration in their acceleration zone. A detailed study of the two types of
acceleration and of their effect on SSC spectra, but in an one-zone model, can be found in \citet{katarz06}.

In the present paper we  investigate further the expected variability within the two-zone model by letting its key parameters vary in time.
While we keep the same assumptions as in KRM, we add also to the photon spectrum radiation coming from the acceleration zone. In
Sect.~\ref{sec:2} we present the basic points of the model and then examine some interesting cases in Sect.~\ref{sec:3}. In Sect.~\ref{sec:4} we
apply our results to \object{1ES\ 1218+304} and discuss them in Sect.~\ref{sec:5}.

\section{The model}
\label{sec:2}

We employ the two-zone acceleration model as developed by KRM. According to this, electrons are accelerated by a non-relativistic shock wave
which propagates along a cylindrical jet and then they escape in a wider region behind the shock where they radiate the bulk of their energy.
Following the KRM notation we shall refer to the region around the advancing shock as the 'acceleration zone' (AZ) and to the escape region as
the 'radiation zone' (RZ). The total size of the source is assumed to have a finite extent.

\subsection{Acceleration zone}

The evolution of the electron distribution function (EDF) in the AZ, $N(\gamma,t)$, is given by the continuity equation
\begin{equation}
\frac{\partial N}{\partial t}+\frac{\partial }{\partial \gamma}(N\dot{\gamma})+\frac{N}{t_{\mathrm{esc}}}=Q\delta(\gamma-\gamma_\mathrm{{inj}}).
\label{eq:az}
\end{equation}
The energy of each particle in this region changes with the rate
\begin{equation}
\dot{\gamma}=\frac{\gamma}{t_{\mathrm{acc}}}-\alpha \gamma^2,
\end{equation}
where the first term describes first order Fermi-type acceleration with the energy independent timescale $t_{\mathrm{acc}}$ and the second
describes the energy losses due to synchrotron radiation in the magnetic field $B$ with
\begin{equation}
\alpha=\frac{\sigma_{\mathrm{T}}B^2}{6\pi m_\mathrm{e}c}.
\end{equation}
Particles are injected in the AZ at some low energy $\gamma_\mathrm{{inj}}m_\mathrm{e}c^2$ with the rate $Q$ and escape from it to the RZ with the energy
independent rate $t_{\mathrm{esc}}^{-1}$.

In our model we assume that variability can be produced by the change of one or more of the parameters $t_{\mathrm{acc}}$, $t_{\mathrm{esc}}$,
$Q$ and $B$ in the AZ and thus treat them as time-dependent in Eq.~(\ref{eq:az}). We define the dimensionless functions that include the
time-dependency through the relations $t_\mathrm{acc}(t)=t_\mathrm{a,0}f_\mathrm{a}(t)$, $t_\mathrm{esc}(t)=t_\mathrm{e,0}f_\mathrm{e}(t)$,
$Q(t)=Q_0f_\mathrm{q}(t)$ and $B(t)=B_\mathrm{0}f_\mathrm{B}(t)$, where $t_\mathrm{a,0}$, $t_\mathrm{e,0}$, $Q_0$ and $B_0$ are the unperturbed
values of the parameters.

The solution of Eq.~(\ref{eq:az}) can be obtained semi-analytically with the method of characteristics. This method requires the solution of the
initial value problem
\begin{equation}
\frac{d\gamma}{dt}=\dot{\gamma}(\gamma,t),\ \gamma(t_*)=\gamma_\mathrm{{inj}},
\end{equation}
which can be written in the form
\begin{equation}
\frac{\varphi_\mathrm{a}(t)}{\gamma}-\frac{\psi(t)}{\gamma_\mathrm{max}}=\frac{\varphi_\mathrm{a}(t_*)}{\gamma_\mathrm{inj}}-\frac{\psi(t_*)}{\gamma_\mathrm{max}},
\label{eq:tst}
\end{equation}
where
\begin{equation}
\gamma_\mathrm{{max}}=(\alpha_0 t_{\mathrm{a,0}})^{-1}
\label{eq:gmax}
\end{equation}
and the functions $\varphi_\mathrm{a}$, $\psi$ are given by the relations
\begin{equation}
\varphi_\mathrm{a}(t)=\exp\left(\int_0^{t/t_\mathrm{a,0}}\frac{dt'}{f_\mathrm{a}(t_\mathrm{a,0}t')}\right) \label{eq:pha}
\end{equation}
and
\begin{equation}
\psi(t)=\int_0^{t/t_\mathrm{a,0}}dt'f_\mathrm{B}^2(t_\mathrm{a,0}t')\varphi_\mathrm{a}(t_\mathrm{a,0}t').
\end{equation}

The solution of Eq.~(\ref{eq:az}) can then be written in the generic closed form
\begin{equation}
N(\gamma,t)=\frac{Q(t_*)\gamma_\mathrm{inj}^2}{\gamma^2\dot{\gamma}(\gamma_\mathrm{inj},t_*)}\frac{\varphi_\mathrm{a}(t)}{\varphi_\mathrm{a}(t_*)}\left(\frac{\varphi_\mathrm{e}(t_*)}{\varphi_\mathrm{e}(t)}\right)^{s-1}S(\gamma;\gamma_\mathrm{inj},\gamma_1(t))
\label{eq:distaz}
\end{equation}
where $s=1+t_{\mathrm{a,0}}/t_{\mathrm{e,0}}$, $S(x;a,b)=1$ if $a\leq x\leq b$ and zero otherwise, and the function $\varphi_\mathrm{e}$ is
given by Eq.~(\ref{eq:pha}) with $f_\mathrm{a}$ replaced by $f_\mathrm{e}$. The quantity $t_*=t_*(\gamma,t)$ in the case where both the
acceleration timescale and the magnetic field are time independent ($f_\mathrm{a}=f_\mathrm{B}=1$) has the analytic expression
\begin{equation}
t_*=t-t_\mathrm{a,0}\ln\left(\frac{\gamma_\mathrm{max}/\gamma_\mathrm{inj}-1}{\gamma_\mathrm{max}/\gamma-1}\right) \label{eq:tst0}
\end{equation}
while in every other case it is obtained from the solution of the implicit Eq.~(\ref{eq:tst}). The upper limit of the EDF is obtained from
Eq.~(\ref{eq:tst}) for $t_*=0$ and it is given by
\begin{equation}
\gamma_1(t)=\frac{\gamma_\mathrm{inj}\varphi_\mathrm{a}(t)}{1+\frac{\gamma_\mathrm{inj}}{\gamma_\mathrm{max}}\psi(t)}.
\end{equation}

In the case where all parameters are time-independent, the unperturbed distribution function of the AZ is
\begin{multline}
N_0(\gamma,t)=Q_0t_\mathrm{a,0}\left(\frac{1}{\gamma_\mathrm{inj}}-\frac{1}{\gamma_\mathrm{max}}\right)^{1-s}\gamma^{-s}\left(1-\frac{\gamma}{\gamma_\mathrm{max}}\right)^{s-2}\\
S(\gamma;\gamma_\mathrm{inj},\gamma_1(t))
\label{eq:EDF}
\end{multline}
and its upper limit is
\begin{equation}
\gamma_1(t)=\frac{\gamma_\mathrm{inj}\mathrm{e}^{t/t_\mathrm{a,0}}}{1+\frac{\gamma_\mathrm{inj}}{\gamma_\mathrm{max}}(\mathrm{e}^{t/t_\mathrm{a,0}}-1)},
\end{equation}
relation valid whenever $f_\mathrm{a}=f_\mathrm{B}=1$. The time it takes for steady-state to establish is a few tens $t_\mathrm{a,0}$ and then
the EDF is a power-law of index $s$ up to the maximum Lorentz factor $\gamma_\mathrm{{max}}$ (Eq.~\ref{eq:gmax}) where acceleration balances
energy losses.

\subsection{Radiation zone}

The escaping particles enter the radiation zone where they radiate a part of their energy in the constant magnetic field $B_0$. Assuming that
synchrotron losses dominate, we can write an equation for the evolution of their differential density $n(\gamma,x,t)$ in the RZ
\begin{equation}
\frac{\partial n}{\partial t}-\frac{\partial}{\partial
\gamma}\left(\alpha_0\gamma^2n\right)=\frac{N(\gamma,t)}{t_\mathrm{esc}}\delta(x-x_\mathrm{sh}(t)),\label{eq:rz}
\end{equation}
where $x_\mathrm{sh}(t)=u_\mathrm{sh}t$ is the position of the shock and $u_\mathrm{sh}$ is the shock speed. The solution of Eq.~(\ref{eq:rz})
is easily found
\begin{equation}
n(\gamma,x,t)=\frac{N(\gamma_*,x/u_\mathrm{sh})}{u_\mathrm{sh}t_\mathrm{esc}(x/u_\mathrm{sh})}\left(\frac{\gamma_*}{\gamma}\right)^2S\left(x;0,x_\mathrm{sh}(t)\right)
\label{eq:distrz}
\end{equation}
with
\begin{equation}
\gamma_*=\left(\frac{1}{\gamma}-\alpha_0(t-x/u_\mathrm{sh})\right)^{-1}.
\end{equation}
Following KRM, we impose the additional restriction $x>x_\mathrm{sh}(t)-L$ on the spatial limits of Eq.~(\ref{eq:distrz}), in order to take into
account the finite size of the RZ, $L$. This is better expressed through the time the shock needs to travel the RZ
\begin{equation}
t_\mathrm{b}=L/u_\mathrm{sh}.
\label{eq:tb}
\end{equation}
Particles not fulfilling this restriction are assumed to escape the RZ in a region of practically zero magnetic field and thus do not contribute
to the total radiation. The EDF in the RZ is then obtained by spatially integrating the function $n(\gamma,x,t)$.

In the case where all parameters are time-independent, the steady-state distribution in the RZ is a broken power-law having the same energy
limits as the EDF in the AZ. The power-law has an index $s$ for energies less than the breaking energy
$\gamma_\mathrm{{br}}\simeq\gamma_\mathrm{{max}}t_{\mathrm{a,0}}/t_\mathrm{b}$ and $(s+1)$ for $\gamma>\gamma_\mathrm{{br}}$. The break is
formed because of the finite size of the source, since the particles with $\gamma<\gamma_\mathrm{br}$ leave the source before they cool.
Radiation from these particles is ignored because of the assumption that the magnetic field declines substantially outside the source. The time
it needs the EDF in the RZ to reach steady-state is of the order of $t_\mathrm{b}$.

\subsection{Radiation spectra}

Knowing the electron distribution functions in both zones, it is easy to compute the emitted synchrotron spectrum in the source frame through
the relation
\begin{multline}
I_\nu(t)=I_\nu^\mathrm{AZ}(t)+I_\nu^\mathrm{RZ}(t)\\=\int_1^\infty d\gamma N(\gamma,t)I_\mathrm{\gamma}(\nu)+\int_1^\infty d\gamma
I_\mathrm{\gamma}(\nu)\int_0^\infty dx n(\gamma,x,t) \label{spectr}
\end{multline}
where $I_\mathrm{\gamma}(\nu)$ is the single particle emissivity for synchrotron radiation that is given by
\begin{equation}
I_\mathrm{\gamma}(\nu)=\frac{\sqrt{3}e^3B}{m_\mathrm{e}c^2}\ F\left(\frac{\nu}{2\gamma^2\nu_0}\right), \label{syngav}
\end{equation}
with
\begin{equation}
F(x)=2x^2\left(K_{4/3}(x)K_{1/3}(x)-\frac{3x}{5}\left(K_{4/3}^2(x)-K_{1/3}^2(x)\right)\right)
\end{equation}
\citep{crusius86} and the characteristic frequency is
\begin{equation}
\nu_0=\frac{3}{4\pi}\frac{eB}{m_\mathrm{e}c}.
\end{equation}
In Eq.~(\ref{spectr}) one can see the two differences that our approach has with the one adopted by KRM. The first is that we calculate
radiation from both zones, while KRM ignore the contribution from the AZ. The second difference is that we do not treat light-travel effects
inside the source. This assumption is equivalent to taking $u_\mathrm{sh}\ll c$ and in that case the shock speed is no longer a parameter of the
problem. Our model thus considers the case of non-relativistic shocks. In the opposite case, light-travel effects can be important depending on
the frequency (see the discussion in KRM) and then the intrinsic variations are smoothed out on the light-crossing timescale. A study of these
phenomena in the internal-shock model for blazars can be found in \citet{boettchder10}.

\section{Flare profiles}
\label{sec:3}

The model that we presented in the previous section can produce a wide variety of flaring behaviors by varying one or more of its basic
parameters. The temporal behavior of these is essentially another free parameter as no theory can provide this information. While in many
applications a short duration impulsive change is assumed \citep[][KRM]{mastikirk97}, here we adopt the Lorentz profile for the changes which has a
pulse shape and analytic expression
\begin{equation}
f_\mathrm{L}(t;t_0,w,n)=1+(n-1)\frac{w^2}{4(t-t_0)^2+w^2}.
\end{equation}
At $t=t_0$ the pulse shows a maximum or a minimum, i.e. $f_\mathrm{L}(t_0;t_0,w,n)=n$, depending on whether $n>1$ or $n<1$, while for $\vert
t-t_0\vert \gg w/2$ it is $f_\mathrm{L}=1$. The quantity $w$ is the full width at half maximum (or minimum, FWHM) of the pulse, since
$f_\mathrm{L}(t_0\pm \frac{w}{2};t_0,w,n)=(n+1)/2$.

\subsection{Change in the injection rate}

A change in the injection rate can be attributed to the encounter of the shock with a higher or lower density area. The increase of the
injection rate leads to an increase in the overall normalization of the EDFs and thus to the production of a flare. The case of a step-function
change was examined in KRM. Here we study it for a Lorentz type change in more detail.

\begin{figure}
\centering \resizebox{\hsize}{!}{\includegraphics{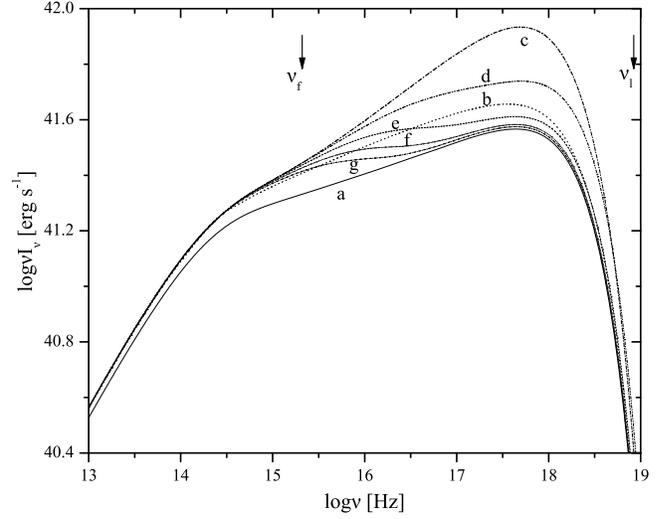}} \caption{Snapshots of the synchrotron spectrum in the source frame. The curve
labeled 'a' is the steady-state spectrum, while labels 'b' to 'g' are the snapshots at $t=t_0+4kt_\mathrm{a,0}$ for $k=2,...,7$ respectively.
The vertical arrows define the frequency range $(\nu_\mathrm{f},\nu_\mathrm{l})$ for which the light-curves of Fig.~\ref{flq0a} are calculated.}
\label{flq0sp}
\end{figure}

\begin{figure}
\centering \resizebox{\hsize}{!}{\includegraphics{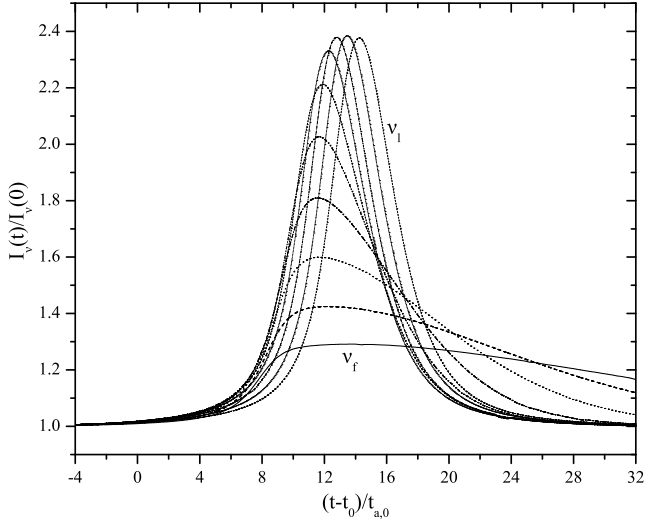}} \caption{Flares that are produced by a Lorentzian change of the injection rate in
the frequency range $\nu_\mathrm{f}=10^{-3}\nu_\mathrm{max}\leq \nu \leq 10^{0.6}\nu_\mathrm{max}=\nu_\mathrm{l}$ with step
$10^{0.4}\nu_\mathrm{max}$ (from bottom to top, the first and the last curve are labeled with their corresponding frequency).} \label{flq0a}
\end{figure}

A few snapshots of the synchrotron spectrum when the injection rate varies according to $f_\mathrm{q}(t)=f_\mathrm{L}(t;t_0,w,n)$ with
$w=3t_\mathrm{a,0}$, $n=3$ and $t_0\gg t_\mathrm{b}$ -- so that steady-state has been reached long before $t=t_0$ -- are shown in
Fig.~\ref{flq0sp}. For the steady-state spectrum, the set of parameters $\gamma_\mathrm{{inj}}=10$, $B=0.5\ \mathrm{G}$,
$t_\mathrm{b}=100t_\mathrm{a,0}$, $Q_0=10^{46}\,\mathrm{s}^{-1}$ and $t_\mathrm{a,0}$, $t_\mathrm{e,0}$ such that $\gamma_\mathrm{{max}}=10^6$
and $s=1.8$, is adopted. The maximum synchrotron frequency in the source frame, $\nu_\mathrm{max}=\gamma_\mathrm{max}^2\nu_0$, is then a few
keV. The corresponding light-curves are shown in Fig.~\ref{flq0a}.

It is evident from Fig.~\ref{flq0a} that the form of the flares is a function of frequency. In order to quantify this, we define for each flare
the following parameters. If $y(t)=I_\nu(t)/I_\nu(0)$ is the form of the light-curve normalized to the steady-state luminosity, $I_\nu(0)$, then
we define the time at which the flare peaks $t_\mathrm{pk}$ and its peak value $y_\mathrm{pk}\equiv y(t_\mathrm{pk})$. We also define the flux
doubling time during the rise of the flare, $t_\mathrm{r}$, and the flux halving time during the decay, $t_\mathrm{d}$, through the relations
$y(t_\mathrm{pk}-t_\mathrm{r})=y(t_\mathrm{pk}+t_\mathrm{d})=(1+y_\mathrm{pk})/2$. The ratio $t_\mathrm{r}/t_\mathrm{d}$ is then a measure of
the time symmetry of the flare and $w_\mathrm{fl}=t_\mathrm{r}+t_\mathrm{d}$ is the FWHM of the flare. We should note here that we do not fit
each flare with a specific function, but rather we calculate the above quantities independently from the exact form of the flares.

\begin{figure}
\centering
\includegraphics[width=0.4\textwidth]{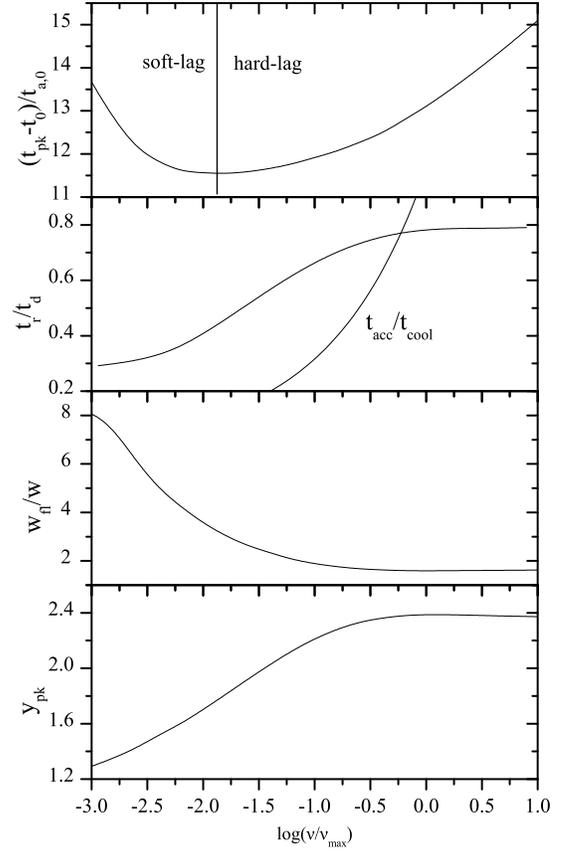}
\caption{Characteristics of the flares shown in Fig.~\ref{flq0a}.} \label{flq0b}
\end{figure}

In Fig.~\ref{flq0b} we plot the quantities $t_\mathrm{pk}$, $t_\mathrm{r}/t_\mathrm{d}$, $w_\mathrm{fl}$ and $y_\mathrm{pk}$ as functions of
frequency. In the first plot for $\nu\lesssim 0.02\nu_\mathrm{max}$ the time at which the flare peaks decreases with frequency, while for
$\nu\gtrsim 0.02\nu_\mathrm{max}$ it increases. This means that for $\nu\lesssim 0.02\nu_\mathrm{max}$ the low-frequency flares precede the
high-frequency ones, or, in other words, the flares show a 'soft-lag'. The opposite happens for $\nu\gtrsim 0.02\nu_\mathrm{max}$ where the
flares show a 'hard-lag'.

As noted in KRM, who also find this result, soft-lags can be attributed to electrons for which the acceleration timescale is much faster than
the cooling timescale, therefore for all practical reasons these can be considered as pre-accelerated or injected at high energies with a ready
power-law \citep[see, for example][]{mastikirk97}. On the other hand, hard-lags are observed at higher frequencies where the two timescales are
comparable, therefore the observer sees the wave of freshly accelerated particles moving to high energies.

Further insight can be gained by considering the exact changes which the electron populations experience. The distribution function of the AZ
can be written in this case as
\begin{equation}
N(\gamma,t)=N_0(\gamma,t)f_\mathrm{L}(t_*;t_0,w,n)
\end{equation}
with $t_*$ given by Eq.~(\ref{eq:tst0}). This result indicates that the initial pulse propagates in the EDF from low to high energies, since its
peak is located at $\gamma=\gamma_1(t-t_0)$, as can be easily found by setting $t_*=t_0$. Moreover, this moving pulse preserves the shape of the
initial pulse and thus it is symmetric around its center and has an amplitude equal to $n$. So, if we consider radiation only from the AZ, we
expect to get roughly symmetric flares of the same amplitude which exhibit hard-lags.

In the radiation zone the situation is more complicated, since the EDF there results from the integration of Eq.~(\ref{eq:distrz}) along the
extent of the source. The hard-lag behavior of the AZ is seen here also, but only in high energies, where the contribution from the particles
residing just before the shock dominates in the integral. In lower energies the perturbation propagates towards the opposite direction, as the
intrinsic variations of $N(\gamma,t)$ are smoothed out by the spatial integration and only the effect of the fading pulse in the AZ stands out.

In the second and third plot of Fig.~\ref{flq0b} one sees that the width of the flares decreases with frequency, while the ratio
$t_\mathrm{r}/t_\mathrm{d}$ increases. This means that the flares become narrower and more symmetric as the frequency increases. If the rise
time of the flares was connected with the acceleration timescale and the decay with the cooling timescale, then the ratio
$t_\mathrm{r}/t_\mathrm{d}$ will be proportional to
\begin{equation}
\frac{t_\mathrm{acc}}{t_\mathrm{cool}}=\frac{t_\mathrm{a,0}}{(\alpha_0\gamma)^{-1}}=\frac{\gamma}{\gamma_\mathrm{max}}\simeq\left(\frac{\nu}{\nu_\mathrm{max}}\right)^{1/2}.
\end{equation}
The plot of this curve in the $t_\mathrm{r}/t_\mathrm{d}$ diagram reveals that this is not the case and thus the relation of $t_\mathrm{r}$,
$t_\mathrm{d}$ to the acceleration and cooling timescales is more complicated.

Finally, in the last plot of Fig.~\ref{flq0b} one sees that the amplitude of the flares increases with frequency until it saturates to a
constant value which is somewhat lower than the amplitude of the electron variation.
The level of saturation depends on the length of the pulse, because the longer the duration of the burst becomes, the closer
the EDF gets to a steady state.

\subsection{Change in the escape timescale}

\begin{figure}
\centering \resizebox{\hsize}{!}{\includegraphics{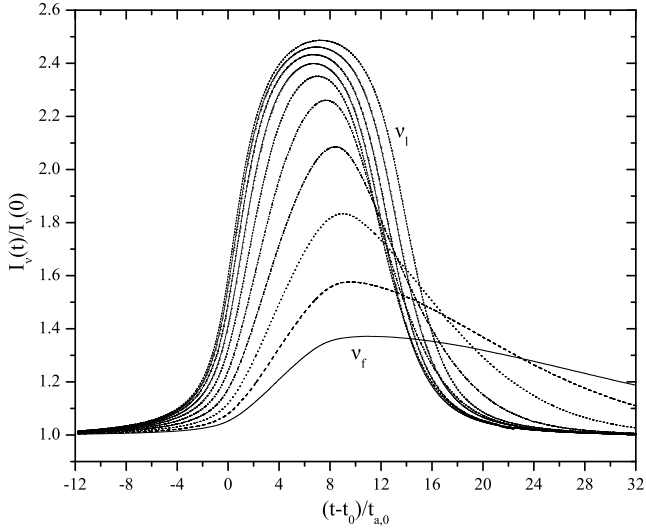}} \caption{Flares that are produced by a Lorentzian change of the escape timescale
in the frequency range $\nu_\mathrm{f}=10^{-3}\nu_\mathrm{max}\leq \nu \leq 10^{0.6}\nu_\mathrm{max}=\nu_\mathrm{l}$ with step
$10^{0.4}\nu_\mathrm{max}$ (from bottom to top, the first and the last curve are labeled with their corresponding frequency).} \label{fltea}
\end{figure}

As a next case we examine a perturbation in the escape timescale. More precisely we assume that some physical process impedes momentarily
particle escape, i.e. decreases the rate at which particles escape from the acceleration region. An inspection of the EDF (Eq.~\ref{eq:EDF})
shows that an increase in the escape timescale leads to a decrease in the electron index. This will produce a flare, the form of which, at
various frequencies, is shown in Fig.~\ref{fltea}. We assume that the change of the escape timescale follows again a Lorentzian variation, i.e.
$f_\mathrm{e}(t)=f_\mathrm{L}(t;t_0,w,n)$ with $w=3t_\mathrm{a,0}$, $n=4/3$ and $t_0\gg t_\mathrm{b}$, while the parameters of the steady-state
spectrum are the same as before.

\begin{figure}
\centering
\includegraphics[width=0.4\textwidth]{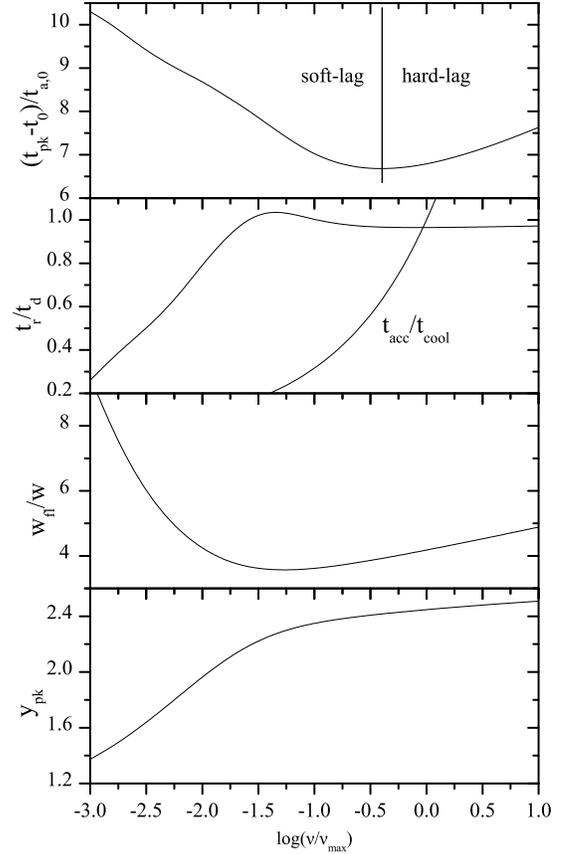}
\caption{Characteristics of the flares shown in Fig.~\ref{fltea}.} \label{flteb}
\end{figure}

The specific features of the flares are shown in Fig.~\ref{flteb}. As before the flares exhibit a soft-lag in low frequencies and a hard-lag
close to the maximum frequency. The frequency range in which hard-lags are observed is much smaller than in the previous case and also the
flares now peak closer to the peak of the perturbation pulse, i.e. closer to $t=t_0$.

The shape of the flares is asymmetric in time at low frequencies, having a rise time which is shorter than the decay one (second plot of
Fig.~\ref{flteb}). This however becomes symmetric, i.e. $t_\mathrm{r}\simeq t_\mathrm{d}$, at higher frequencies. The rising part of the curve
indicates that $t_\mathrm{r}/t_\mathrm{d}$ is related to $t_\mathrm{acc}/t_\mathrm{cool}$ and thus the rise of the flares is controlled by the
acceleration and the decay by the cooling timescale.

The width of the low-frequency flares decreases as in the previous case but in higher frequencies it increases. We note, however, that
$w_\mathrm{fl}\gg w$, i.e. even a narrow pulse in the escape timescale produces a rather broad X-ray flare. The amplitude of the flares
increases with frequency as in the previous case, but now it does not saturate to a maximum value.

The EDF of the AZ, Eq.~(\ref{eq:distaz}), reads in this case
\begin{multline}
N(\gamma,t)=N_0(\gamma,t) \exp \left[\frac{w(n-1)(s-1)}{2\sqrt{n}}\times\right.\\
\left.\left(\tan^{-1} \left(2\frac{t-t_0}{w\sqrt{n}}\right)-\tan^{-1} \left(2\frac{t_{*}-t_0}{w\sqrt{n}} \right) \right) \right]
\end{multline}
where $t_*$ is given by Eq.~(\ref{eq:tst0}). The perturbation in the EDF propagates again from low to high energies but is no longer
pulse-shaped and has a much broader extent. Moreover, the perturbation moves twice as fast now, since the location of its peak is given by
$\gamma=\gamma_1(2(t-t_0))$. One notices that, qualitatively speaking, the trends in the present case are similar to the previous one.
There is a simple explanation for this. The perturbation introduced causes the escape timescale to increase first and then decrease back to its
original value. This in turn causes the electron spectral index to flatten and then steepen again. Therefore, the number of
accelerated particles increases and then decreases to its unperturbed value. Thus this case can be considered as equivalent to the previous one
which treated variations in the injection of particles at low energies. Note that, in the present case, the flattening of the EDF is partly
compensated from the fact that, at the same time, the particles decrease in the RZ (see Eq.~\ref{eq:distrz}).

\subsection{Change in the acceleration timescale}

\begin{figure}
\centering \resizebox{\hsize}{!}{\includegraphics{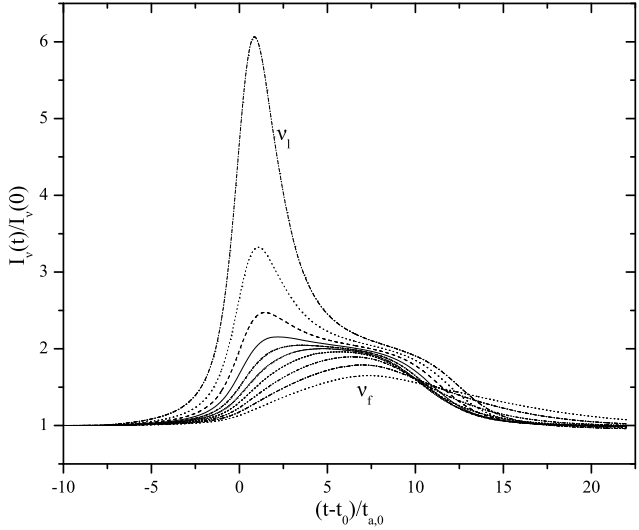}} \caption{Flares that are produced by a Lorentzian change of the acceleration
timescale in the frequency range $\nu_\mathrm{f}=10^{-2.1}\nu_\mathrm{max}\leq \nu \leq 10^{0.6}\nu_\mathrm{max}=\nu_\mathrm{l}$ with step
$10^{0.3}\nu_\mathrm{max}$ (from bottom to top, the first and the last curve are labeled with their corresponding frequency).} \label{flta}
\end{figure}

As a next example we investigate changes in the acceleration timescale and more specifically the variations induced when this decreases. This
change has two effects on the EDFs and, therefore, to the produced spectrum. First, an inspection of the EDF (Eq.~\ref{eq:EDF}) reveals that if
the acceleration timescale decreases, then the electron index becomes flatter. Second, the maximum electron energy will increase -- see
Eq.~(\ref{eq:gmax}). The flares which are produced for a Lorentzian change with $w=3t_\mathrm{a,0}$ and $n=3/4$ are shown in Fig.~\ref{flta}.

As frequency increases, the flare shapes become more asymmetric and peak at earlier times. In this case we get only soft-lag flares. This can
easily be explained since the increase of $\gamma_\mathrm{max}$ in the EDFs leads to the simultaneous increase of the maximum emitted frequency
in the spectrum. This produces a large flare at high frequencies, since during pre-flare this part of the spectrum was in the exponentially
decaying synchrotron regime.

\subsection{Change in the magnetic field}

\begin{figure}
\centering \resizebox{\hsize}{!}{\includegraphics{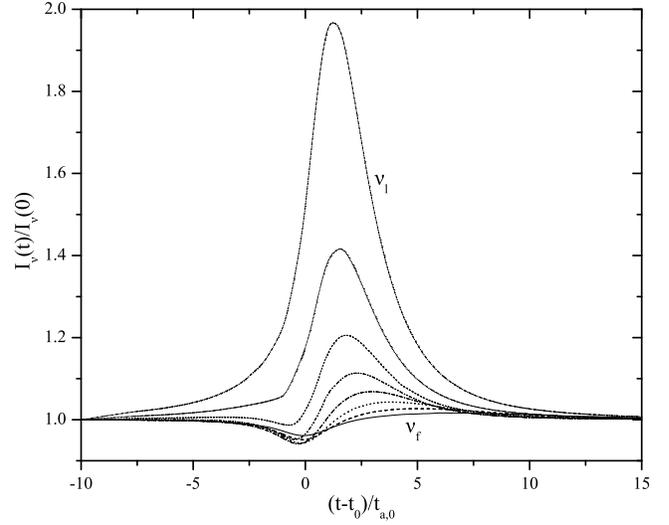}} \caption{Flares that are produced by a Lorentzian change of the magnetic field in
the frequency range $\nu_\mathrm{f}=10^{-1.9}\nu_\mathrm{max}\leq \nu \leq 10^{0.2}\nu_\mathrm{max}=\nu_\mathrm{l}$ with step
$10^{0.3}\nu_\mathrm{max}$ (from bottom to top, the first and the last curve are labeled with their corresponding frequency).} \label{flb}
\end{figure}

As a last case we examine variations of the magnetic field strength which are assumed to occur in the AZ only. A decrease in the magnetic field
leads to an increase of the maximum electron energy. Note that the maximum photon frequency increases also, even if $B$ decreases and this
results to a production of a flare. The form of the flares which are produced for a Lorentzian change of the magnetic field with
$w=3t_\mathrm{a,0}$ and $n=0.8$ is shown in Fig.~\ref{flb}. A new feature that appears is a small decrease of the flux preceding the low
frequency flares. This is because the initial drop of the magnetic field leads to the decrease of the total power radiated by the electrons and
this is seen in low frequencies since the high-frequency spectrum rises due to the increase of $\gamma_\mathrm{max}$ which compensates for this
reduction.

The flares exhibit only soft-lags for a similar reason to that of the previous case. The flares also become narrower and more symmetric with
frequency. In total, this case has many similarities with the change of the acceleration timescale, since in both cases the maximum electron
energy follows the same temporal behavior.

\section{Application to \object{1ES\ 1218+304}}
\label{sec:4}

\object{1ES\ 1218+304} is a high-frequency peaked BL Lac object at a redshift $z=0.182$. It was observed in May 2006 with Suzaku and the
analysis of the observations \citep{sato08} revealed that the flares in the various X-ray energy bands exhibit a hard-lag. Moreover, the flare
shapes are asymmetric in time and their amplitude becomes larger with photon energy. These features resemble the ones found in the previous
section for the cases of change of the injection rate or the escape timescale.

\begin{figure}
\centering
\includegraphics{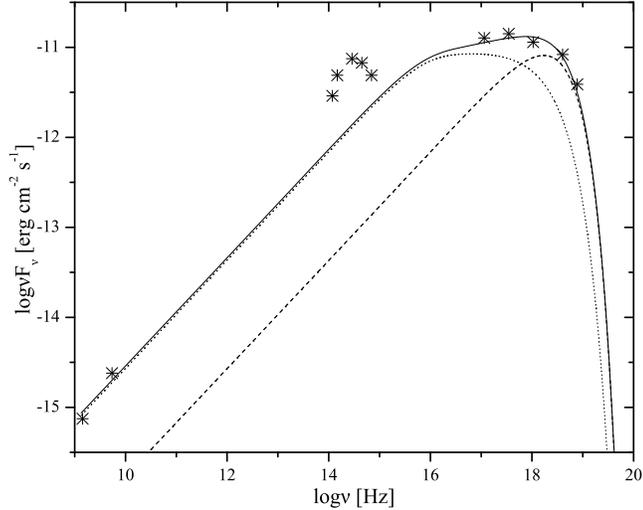}
\caption{Radio to X-rays spectrum of \object{1ES\ 1218+304} (starred symbols, data taken from \citealt{sato08}) together with the steady-state
theoretical spectrum (solid line) for the parameters given in text. The contributions from the acceleration zone (dashed line) and from the
radiation zone (dotted line) are also shown. The optical data can be attributed to the host galaxy of \object{1ES\ 1218+304} \citep{ruger10} and
are thus not included in the fit.}\label{spec}
\end{figure}

The radio to X-rays spectrum of \object{1ES\ 1218+304} is shown in Fig.~\ref{spec} together with the steady-state spectrum of our model as this
is given in Eq.~(\ref{spectr}). The transformation to the observer's frame is made through the relations
\begin{align}
F_{\nu}^{\mathrm{obs}} &= \delta^3(1+z)\frac{I_\nu}{4\pi d_\mathrm{L}^2} \\
    \nu_{\mathrm{obs}} &= \frac{\delta}{1+z}\nu
\end{align}
where $\delta$ is the Doppler factor and $d_\mathrm{L}=880\ \mathrm{Mpc}$ is the luminosity distance of \object{1ES\ 1218+304}, for the flat
universe with cosmological parameters $\Omega_\Lambda=0.7$, $\Omega_\mathrm{m}=0.3$ and $H_0=70\ \mathrm{km}\ \mathrm{s}^{-1}\
\mathrm{Mpc}^{-1}$.

The maximum photon frequency in the spectrum is $\nu_\mathrm{max}^\mathrm{obs}\simeq 3.6\ 10^{18}\ \mathrm{Hz}$, leading to the relation
\begin{equation}
\delta B\gamma_\mathrm{max}^2\simeq 10^{12}\mathrm{G}. \label{eq:fit}
\end{equation}
We choose a random combination of these parameters so that Eq.~(\ref{eq:fit}) is satisfied and also take $\gamma_\mathrm{inj}=10$,
$t_\mathrm{b}=20t_\mathrm{a,0}$ and $s=1.8$. The value of $t_\mathrm{b}$ is dictated by the frequency where the spectrum appears to break,
$\nu_\mathrm{br}^\mathrm{obs}\simeq 9\ 10^{15}\ \mathrm{Hz}$, while the value of the electron index is chosen so that the photon index in radio
frequencies is $\alpha_\mathrm{r}\simeq 0.6$. The parameter $\gamma_\mathrm{inj}$ has a small effect in the normalization of the spectrum and
thus its value is unimportant.

\begin{figure}
\centering
\includegraphics[width=0.45\textwidth]{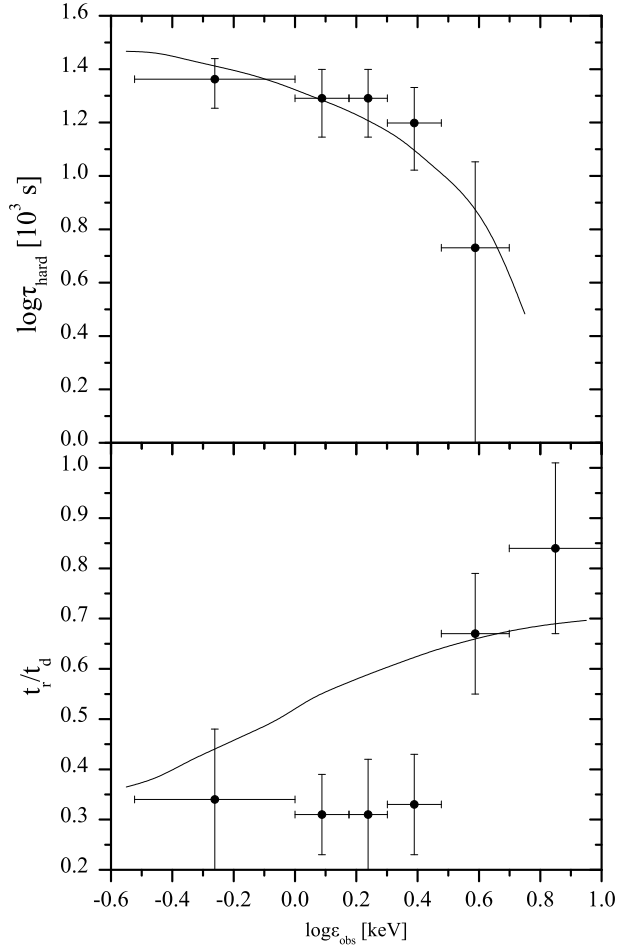}
\caption{Amount of hard-lag and the ratio $t_\mathrm{r}/t_\mathrm{d}$ of \object{1ES\ 1218+304} in the various X-ray energy bands together with
the corresponding theoretical curves.}\label{onees}
\end{figure}

The remaining parameters can be further constrained by the variability timescale of \object{1ES\ 1218+304}. From the values of
$t_\mathrm{pk}(\varepsilon)$ given in \citeauthor{sato08} for the time at which the flares peak, we deduce that the variation responsible for
producing these flares is the change in the injection rate. This is because the energy range in which $t_\mathrm{pk}$ is rising is quite broad
and rules out the case of the escape timescale change. As in \citeauthor{sato08} we define the amount of hard-lag simply by the difference of
the peak times of the flares in the various energy bands from the peak time of the flare in the highest energy band, namely
\begin{equation}
\tau_\mathrm{hard}(\varepsilon)=t_\mathrm{pk}(\varepsilon_\mathrm{max})-t_\mathrm{pk}(\varepsilon),
\end{equation}
where $\varepsilon_\mathrm{max}=\sqrt{50}\ \mathrm{keV}$ is the logarithmic mean energy of the $5-10\ \mathrm{keV}$ energy band. The theoretical
curve for $\tau_\mathrm{hard}(\varepsilon)$ for the Lorentzian change of the injection rate with parameters $w=1.7t_\mathrm{a,0}$ and $n=1.4$,
is shown in the upper panel of Fig.~\ref{onees} together with the observational data. From the fit we estimate the acceleration timescale
$t_\mathrm{a,0}=1.9\ 10^5\ \delta_{10}\ \mathrm{s}$, where $\delta_{10}=\delta/10$. This relation when combined with Eq.~(\ref{eq:fit}) enables
us to express the magnetic field and maximum electron energy of \object{1ES\ 1218+304} as $B=0.06\ \delta_{10}^{-1/3}\ \mathrm{G}$ and
$\gamma_\mathrm{max}=1.3\ 10^6\ \delta_{10}^{-1/3}$.

Since the Doppler factor cannot be determined, we choose $\delta_{10}=1$ and the unperturbed value of the injection rate is then $Q_0=3.2\
10^{45}\ \mathrm{s}^{-1}$. With these parameters we can estimate the size of the source $L= 1.1\ 10^{16}(u_\mathrm{sh}/0.1c)\ \mathrm{cm}$ and
the energy content of the source in electrons as $\mathcal{E}_\mathrm{e}=3\ 10^{48}\, \mathrm{erg}$.

In the lower panel of Fig.~\ref{onees} we plot the theoretical curve of the asymmetry of the flares, $t_\mathrm{r}/t_\mathrm{d}$, together with
the observational data. Our definition of $t_\mathrm{r}$ and $t_\mathrm{d}$ differs from the one in \citeauthor{sato08} but the ratio of the two
quantities is the same in both cases. One sees that we reproduce the general trend of the data quite well but we overestimate the ratio
$t_\mathrm{r}/t_\mathrm{d}$ in the energy interval $1-3\ \mathrm{keV}$.

\section{Conclusions}
\label{sec:5}

In the present paper we have examined the expected variability of blazars in the context of the two-zone acceleration model. By varying four of
the key parameters of the model, we found that in all cases flaring can be induced. In all cases the simulated flares exhibit soft lags -- only
in two cases, i.e. when varying the injection rate of particles or their escape rate, one can find frequency regimes where hard lags appear.
Thus, we find that this scheme cannot produce hard lags at frequencies much lower than the maximum synchrotron frequency. Another conclusion is
that the flares tend to mimic the electron variations at high frequencies. As a rule at low frequencies flares tend to be asymmetric with rise
timescales much faster than decay ones.

We were able also to make an acceptable fit to the hard lag flare of \object{1ES\ 1218+304} \citep{sato08} by varying the  rate of particles
injected into the acceleration process. We note that the fit can be improved if we were to vary two parameters instead of one but this lies
outside of the aims of the present paper. The fitting we obtained gives parameters well within the ones accepted from blazar modeling. Note that
the acceleration timescale deduced ($t_\mathrm{a,0}\simeq2~10^5\delta_{10}$s) might be long for first order Fermi acceleration
\citep{tammiduffy09}, however we have to emphasize that from the setup of the model, both acceleration and escape timescales were assumed to be
independent of  the energy $\gamma$ and the magnetic field strength $B$, therefore we do not try to connect them directly with theories of
particle acceleration \citep{drury83,blandfordeichler87}.

We have restricted our analysis to the cases where synchrotron losses dominate and did not consider the inverse Compton scattering as a possible
electron energy loss mechanism. This can be justified for sources where the energy density due to magnetic fields exceeds the one due to
photons. In this case the variability patterns due to inverse Compton scattering will, in general, follow the ones due to synchrotron
\citep{km99}. On the other hand, an example of a model dealing with X- and gamma-ray variability where inverse Compton losses might be of
importance can be found in \citet{mm08} -- note, however, that this is an one-zone acceleration model.

Our model, despite its limitations, is capable of producing a wide variety of flaring behaviors that could, in principle, be tested against the
growing X-ray data on blazar variability. We note that recently, \citet{garsonetal10} have also attributed the properties of X-ray flares on
intrinsic changes of the acceleration process.

\begin{acknowledgements}
The authors would like to thank the anonymous referee for his/her comments that helped improve the manuscript. KM acknowledges financial support
from the Greek State Scholarships Foundation (IKY).
\end{acknowledgements}

\bibliographystyle{aa}
\bibliography{refs}

\end{document}